\documentclass[11pt,a4paper]{article}
\usepackage{latexsym,amsthm,amssymb}
\usepackage{booktabs,threeparttable}

\newcommand{\be}{\begin{equation}}
\newcommand{\ee}{\end{equation}}
\newcommand{\bea}{\begin{eqnarray}}
\newcommand{\eea}{\end{eqnarray}}
\newcommand{\ba}{\begin{array}}
\newcommand{\ea}{\end{array}}
\newcommand{\bean}{\begin{eqnarray*}}
\newcommand{\eean}{\end{eqnarray*}}
\newcommand{\no}{\nonumber}
\newcommand{\al}{\alpha}

\newcommand{\Ga}{\Gamma}
\newcommand{\ga}{\gamma}

\newcommand{\la}{\lambda}

\newcommand{\geqs}{\geqslant}
\newcommand{\mL}{{\mathcal L\/}}

\newcommand{\mM}{{\mathcal M\/}}

\newcommand{\mF}{{\mathcal F\/}}

\newcommand{\bet}{\beta}
\newcommand{\de}{\delta}
\newcommand{\ep}{\epsilon}
\newcommand{\sig}{\sigma}

\newcommand{\pa}{\partial}

\newcommand{\lan}{\langle}
\newcommand{\ran}{\rangle}

\begin{document}
\title
     {\bf A Note on Symmetries of WDVV Equations\/}
\author
{ Yu-Tung Chen$^1$, Niann-Chern Lee$^2$ and Ming-Hsien Tu$^3$
 \footnote{E-mail: phymhtu@ccu.edu.tw} \\ \\
  $^1$
  {\it Department of Computer Science, National Defense University, Tauyuan, Taiwan\/},\\
    $^2$
{\it General Education Center, National Chin-Yi University of Technology, Taichung, Taiwan\/},\\
   $^3$
  {\it Department of Physics, National Chung Cheng University, Chiayi, Taiwan\/}\/}
\date{\today}
  \maketitle
  \begin{abstract}
  We investigate symmetries of Witten-Dijkgraaf-E.Verlinde-H.Verlinde (WDVV)
   equations proposed by Dubrovin from bi-hamiltonian
  point of view. These symmetries can be viewed as canonical Miura transformations
  between genus-zero bi-hamiltonian systems of hydrodynamic type. In particular,
  we show that the moduli space of two-primary models under symmetries of WDVV can
  be characterized by the polytropic exponent $h$. Furthermore, we also discuss the
  transformation properties of  free energy at genus-one level.
\end{abstract}

\newpage
\section{Introduction}
A Frobenius manifold is a kind of complex manifold $\mM$ whose
tangent space locally defines a commutative and associative algebra with a unit element.
The notion of Frobenius manifolds is related to many subjects in mathematics and theoretical
physics, such as integrable systems, quantum cohomology, symplectic geometry, singularity theory,
topological field theories and mirror symmetry etc.
(see e.g. \cite{Kr92,Dub96,Man99,Hit97,Her02,HKK03} and references therein).
 One way to define an $N$-dimensional Frobenius manifold $\mM$ is to construct a
 quasi-homogeneous function  $F(t)$ of $N$ variables $t=(t^1, t^2, \ldots, t^N)$
  such that the associated functions,
\begin{equation}
 c_{\alpha\beta\gamma}(t)= \frac{\pa^3 F(t)}{ \pa t^\alpha \pa t^\beta \pa t^\gamma},
 \quad \alpha,\beta,\gamma=1,\ldots,N
 \label{def-c}
\end{equation}
satisfy the following conditions\cite{Dub96} :
\begin{itemize}
\item
 The metric $\eta_{\alpha\beta}=c_{1\alpha\beta}$ is
 constant and  non-degenerate (for the discussion of degenerate cases, see \cite{Str99}).
\item
 The functions
 $c_{\beta\gamma}^{\alpha}=\eta^{\alpha\sigma}c_{\sigma\beta\gamma}$ with
$\eta^{\alpha\beta}=(\eta_{\alpha\beta})^{-1}$
  define the structure constants of an associative and commutative algebra
 ${\mathbf A\/}_t$ of dimension $N$ so that the basis $\{e_1,\ldots,e_N\}$:
 $e_{\alpha}\cdot e_{\beta} = c_{\alpha\beta}^{\gamma}e_{\gamma}$,
 with a unity element $e_1$ for all algebra ${\mathbf A\/}_t$ and
 $c_{1\alpha}^{\beta}(t)=\delta_{\alpha}^{\beta}$.
 The associativity condition,
 $(e_\alpha\cdot e_\beta)\cdot e_\ga=e_\alpha\cdot (e_\beta\cdot e_\ga)$ yields
$c_{\alpha \beta}^{\mu}c_{\mu \ga}^{\sigma}=c_{\alpha \ga}^{\mu}c_{\mu \beta}^{\sigma}$
or, by virtue of (\ref{def-c}), a system of non-linear partial differential equations for $F(t)$
\begin{equation}
 \frac{\pa^3 F(t)}{\pa t^{\alpha} \pa t^{\beta} \pa t^{\lambda}} \eta^{\lambda \mu}
 \frac{\pa^3 F(t)}{\pa t^{\mu} \pa t^{\ga}
 \pa t^{\sigma}} = \frac{\pa^3 F(t)}{\pa t^{\alpha} \pa t^{\ga} \pa t^{\lambda}}
 \eta^{\lambda \mu} \frac{\pa^3 F(t)}{\pa
 t^{\mu} \pa t^{\beta} \pa t^{\sigma}}.
 \label{WDVV}
\end{equation}
 \item
 The function $F(t)$ (also called primary free energy) satisfies a quasi-homogeneity
 condition,
\be
 {\mathcal L\/}_{E} F= d_{F}F+ (\mbox{quadratic~terms}),
 \label{quasi}
\ee
where $ E(t)=\sum_{\alpha}(d_\alpha t^{\alpha}+r^\alpha)\pa_\alpha=\sum_\alpha E^\alpha\pa_\alpha$
 is known as the Euler vector field.
\end{itemize}
The associativity equation (\ref{WDVV}) together with quasi-homogeneity  condition
(\ref{quasi}) constitute  the so-called Witten-Dijkgraaf-E.Verlinde-H.Verlinde (WDVV)
equations\cite{Wit90,DVV91}.
It is convenient to set $d_1=1$ and introduce new parameters
$q_1=0, q_2,\cdots, q_N=d$ with $q_\alpha+q_{N-\alpha+1}=d$ in such a way that
$q_\alpha=1-d_\alpha$ and $ d=3-d_F$.

Given any solution (or primary free energy) of the WDVV equations,
one can construct a Frobenius manifold ${\mathcal M\/}$ associated with it
and define a unique structure of a Frobenius algebra $({\mathbf A\/}_t, \lan,\ran)$
on the tangent plane $T_t\mM$  such that
$\lan\pa_\alpha\cdot\pa_\beta, \pa_\ga\ran=\pa_\alpha\pa_\beta\pa_\ga F(t),
\lan\pa_\alpha,\pa_\beta\ran=\eta_{\alpha\beta}$
 with $\pa_\alpha\cdot\pa_\beta=c_{\alpha\beta}^\ga(t)\pa_\ga,
c_{\alpha\beta}^\ga=\eta^{\ga\de}\pa_\alpha\pa_\beta\pa_\de F(t)$
     where we set $\pa_1$ the unity of the algebra.
   On such a manifold one may interpret $\eta^{\alpha\beta}=(\eta_{\alpha\beta})^{-1}$
   as a flat metric and $t^{\alpha}$ the flat coordinates.

The solution space of the WDVV equations is quite rich.
It can be shown \cite{Dub92} that if $\eta_{11}=0$
and all $d_\alpha$ are distinct, then by a properly change of variables $t^\alpha$, the
matrix $\eta_{\alpha\beta}$ has an anti-diagonal form
$ \eta_{\alpha\beta}=\delta_{\alpha+\beta,N+1}$ and
\[
 F(t)=\frac{1}{2}(t^1)^2t^N+
 \frac{1}{2}t^1\sum_{\alpha=2}^{N-1}t^{\alpha}t^{N-\alpha+1}+f(t^2,\ldots,t^N).
\]
The first nontrivial example is the case for $N=2$, where $F(t)=\frac{1}{2}(t^1)^2t^2+f(t^2)$
and $E=t^1\pa_1+(1-d)t^2\pa_2$.
The associativity equation (\ref{WDVV}) are empty, while the quasi-homogeneity conditions
(\ref{quasi}) yields  the equation,
 $t^2f'(t^2)=s f(t^2)+k_1+k_2t^2+k_3(t^2)^2, s=(3-d)/(1-d)$.
After integrating over $t^2$, we obtain
\[
f(t^2)=(t^2)^s\left(k_1\int^{t^2}\frac{dz}{z^{s+1}}+k_2\int^{t^2}\frac{dz}{z^{s}}+
k_3\int^{t^2}\frac{dz}{z^{s-1}}+k_4\right).
\]
For $d\neq 3, \pm 1$, $f\sim (t^2)^s$ (modulo the quadratic part). For $d=3$,
$f\sim \log t^2$. For $d=-1$, $f\sim (t^2)^2\log t^2$. For $d=1$, the Euler
vector field become $E=t^1\pa_1+2\pa_2$ and $f\sim e^{t^2}$. (see Table \ref{two-free}).
They correspond to two-primary solutions of WDVV equations, in which
 two of  them have logarithmic-type primary free energy.
It turns out that the associated integrable structures behind these two-primary models
are polytropic gas dynamics \cite{Wh74,ON88},
dispersionless Harry Dym (dDym) hierarchy\cite{CT03a}, Benney hierarchy \cite{CT01},  and
dispersionless Toda (dToda) hierarchy\cite{EY94}, respectively.
\begin{center}
\begin{threeparttable}
\caption{Models with two primary fields\cite{Dub96}}
\label{two-free}
\index{G-f@$G$-function}
\begin{tabular*}{140mm}[]{ccccc@{}l}
\toprule
 model &  $F(t)$ &  $E(t)$  & $d$    \\
\midrule
  polytropic gas  &  $\frac{1}{2}(t^1)^2t^2+c_h(t^2)^{h+1}$ &
  $t^1\frac{\pa}{\pa t^1}+\frac{2}{h}t^2\frac{\pa}{\pa t^2}$&
 $1-\frac{2}{h}(h\neq -1, 0, 1)$ \\ \\
    dDym  & $\frac{1}{2}(t^1)^2t^2-\frac{1}{2}\log t^2$ &  $t^1\frac{\pa}{\pa t^1}-2t^2\frac{\pa}{\pa t^2}$ &
    3\\ \\
   Benney & $\frac{1}{2}(t^1)^2t^2+\frac{1}{2}(t^2)^2\log (t^2-\frac{3}{2})$  &
   $t^1\frac{\pa}{\pa t^1}+2t^2\frac{\pa}{\pa t^2}$&
    -1\\ \\
   dToda & $\frac{1}{2}(t^1)^2t^2+e^{t^2}$ &  $t^1\frac{\pa}{\pa t^1}+2\frac{\pa}{\pa t^2}$  & 1\\
  \bottomrule
\end{tabular*}
\end{threeparttable}
\end{center}
For $N=3$, solutions of WDVV can be reducible to a particular case of the
Painlev\'e-VI equation\cite{Dub96}.
For $N>3$, since the over-determined system (\ref{WDVV}) are more complicated,
it is not an easy task  to classify its associated integrable systems completely\cite{DZ01}.

In\cite{Dub96}, Dubrovin introduced two kind of nontrivial symmetries of  WDVV equations, which
seems to provide another point of view to study the integrability associated with
Frobenius manifolds. A symmetry  of the WDVV equations  is the transformation
\bean
t^\alpha&\mapsto&\hat{t}^\alpha,\no\\
\eta_{\alpha\beta}&\mapsto&\hat{\eta}_{\alpha\beta},\no\\
F(t)&\mapsto&\hat{F}(\hat{t})
\eean
preserving the equations. Some examples have been given  to demonstrate
symmetries of WDVV in \cite{Dub96}.
However, even for the simplest case ($N=2$),
it still lacks a general discussion to the solution space of WDVV equations using this symmetries.
In this work we try to explore the solution space of two-primary models from symmetry point of view,
since  symmetry is of central importance  to the classification
programme for the integrable structures of Frobenius manifolds\cite{DZ01}.
We shall show that symmetries of WDVV provide Miura transformation between
bi-hamiltonian systems of hydrodynamic type associated with Frobenius manifolds.
In particular, we verify the canonical property of symmetries of WDVV and figure out the moduli
space of the solution space for two-primary models.
More recently,  a  dual formulation of Frobenius
manifolds was introduced by Dubrovin \cite{Dub04}. Although it would be interesting to
study symmetries between dual Frobenius manifolds\cite{RS}, however, we shall not
cover this part in the present work.

This paper is organized as follows. In section 2, we recall some basic
concepts for constructing bi-hamiltonian systems of hydrodynamic type from Frobenius manifolds.
The integrable systems associated with two-primary models  are presented. In section 3,
we show that symmetries of WDVV can be viewed as canonical Miura transformations
 for two-primary models. Moreover,
we find that, under symmetries of WDVV, the moduli space of two-primary models can be
parameterized by a polytropic exponent $h$. Finally, in section 4, we explore the possibility for
promoting symmetries of WDVV to genus-one level.

\section{Bi-hamiltonian Structure}
Dispersionless integrable hierarchies (see e.g.  \cite{TT95,AK96,Li99}) with finite number of
variables are closely related to Frobenius manifolds.
For those bi-hamiltonian hydrodynamic equations \cite{DN84},
a differential-geometric interpretation to  bi-hamiltonian structures can be achieved.
More precisely, writing the Poisson brackets of a bi-hamiltonian hydrodynamic system as
\[
 \{t^{\al}(x), t^{\beta}(y)\}_i = J_i^{\alpha\beta}\delta(x-y),\quad i=1,2;\al,\bet=1,\cdots,N,
 \label{bi-Poi}
\]
where
\be
J_1^{\alpha\beta}(t)=\eta^{\alpha\beta}(t)\pa_x+\ga^{\alpha\beta}_{\sigma}(t)t^\sigma_x,\qquad
J_2^{\alpha\beta}(t)=g^{\alpha\beta}(t)\pa_x+\Ga^{\alpha\beta}_{\sigma}(t)t^\sigma_x,
\label{J12}
\ee
then $\ga_\sigma^{\alpha\beta}(t)$ and $\Ga_\sigma^{\alpha\beta}(t)$ are the contravariant
Levi-Civita connections of the contravariant flat metrics $\eta^{\alpha\beta}(t)$ and
$g^{\alpha\beta}(t)$, respectively. When $\eta^{\alpha\beta}(t)$ is a constant flat metric
(i.e. $\ga^{\alpha\beta}_{\sigma}(t)=0$) we call $t^\alpha$ the flat coordinates.
Given a  primary free energy $F(t)$ the associated bi-hamiltonian structure (\ref{J12})
 can be constructed in the context of Frobenius manifolds \cite{Dub92,Dub96,DZ98}.
 Let us briefly recall the construction.
 Denoting the multiplication of the algebra as $u \cdot v$,
   then another flat metric on ${\mathcal M\/}$ can be defined by
\be
  g^{\al\beta}(t) = E(dt^\al\cdot dt^\beta)=E^\gamma(t) c^{\alpha\beta}_\gamma(t),
\label{sec}
\ee
where $dt^\al\cdot dt^\beta=c_{\ga}^{\al\beta}dt^{\ga} =
\eta^{\al\sigma}c_{\sigma\ga}^{\beta}dt^{\ga}$.
This metric together with the original metric $\eta^{\al\beta}$
define a flat pencil, i.e., $\eta^{\al\beta}+\la g^{\al\beta}$ is flat
as well, and the associated Levi-Civita connection is given by
$\gamma_\sigma^{\alpha\beta}+\la\Gamma_\sigma^{\alpha\beta}$ for any value of
 $\lambda$, which corresponds to the compatible condition in integrable systems.
We remark that, in terms of flat coordinates, the contravariant components of the
Levi-Civita connection associated with  the metric $g^{\alpha\beta}(t)$
can be expressed as\cite{Dub96}
\be
\Ga^{\alpha\beta}_\gamma(t)=c^{\alpha\ep}_\gamma(t)\left(\frac{1}{2}-\mu\right)_\ep^\beta,
\label{Ga-c}
\ee
where the matrix $\mu=diag(\mu_1,\cdots,\mu_N)$ with $\mu_\alpha=q_\alpha-d/2$.
The associated  genus-zero  hierarchy flows can be written as
\[
\frac{\pa t^{\al}}{\pa T^{\beta,n}}
= \{t^{\al}(x), H_{\beta,n}\}_1,
\label{flow}
\]
where the Hamiltonians
\[
 H_{\beta,n} = \int h_{\beta}^{(n+1)}(t(x)) dx
\]
are defined  by the recursive relations \cite{Dub92,Dub96}
 \[
  \frac{\pa^2 h_{\alpha}^{(n+1)}}{\pa t^\beta \pa
t^\gamma}=c_{\beta\gamma}^\sigma \frac{\pa h_{\alpha}^{(n)}}{\pa t^\sigma},\qquad
h_{\alpha}^{(0)}=\eta_{\alpha \beta}t^{\beta},
 \label{rec}
 \]
 and
\[
\pa_Eh_{\beta}^{(n)}
= (n+1-d/2+\mu_{\beta})h_{\beta}^{(n)} + \sum_{k=1}^n(R_k)_\beta^\gamma h_\gamma^{(n-k)}
\label{rec2}
\]
with the constant matrices $R_k$ satisfying
\[
[\mu, R_k]=kR_k,\quad (R_k)_\alpha^\gamma\eta_{\gamma\beta}
=(-1)^{k+1}(R_k)_\beta^\gamma\eta_{\gamma\alpha}.
\]
It is straightforward to show that $\pa t^\alpha/\pa T^{1,0}=\pa t^\alpha/\pa x$
and thus we identify $T^{1,0}=x$.
In fact, one can define the second hamiltonian structure by the recursive relation
\[
\{t^{\al}(x), H_{\beta,n-1}\}_2=(n+\mu_\beta+1/2)\{t^{\al}(x), H_{\beta,n}\}_1+
\sum_{k=1}^n(R_k)_\beta^\gamma\{t^{\al}(x), H_{\gamma,n-k}\}_1,
\]
so that the hierarchy flows can be expressed in a bi-hamiltonian form
\be
\frac{\pa t^{\al}}{\pa T^{\beta,n}}
= \{t^{\al}(x), H_{\beta,n}\}_1=\{t^{\al}(x), \widetilde{H}_{\beta,n-1}\}_2
\label{bi-flow}
\ee
with
\[
\widetilde{H}_{\beta,n-1}=\sum_{k,l}(-1)^k(R_{n-l,k})^\ga_\beta
\frac{H_{\ga,l-1}}{(n+\mu_\beta+1/2)^{k+1}},
\label{newH}
\]
where $R_{0,0}=1, R_{n>0,0}=0$, and $R_{n,l>0}=\sum_{i_1+\cdots +i_l=n}R_{i_1}\cdots R_{i_l}$.
It may be noticed \cite{Dub92} that the expression (\ref{bi-flow}) of commuting flows fails to
satisfy the second structure for the pair $(\beta,n)$ provided that $n+\mu_\beta+1/2=0$.
A Frobenius manifold is resonant if it has such a pair $(\beta,n)$.

Before closing this subsection, we remark that the time parameters $\{T^{n,\alpha}\}$
of hierarchy flows correspond to coupling constants with respect to the operators
$\sigma_{n,\alpha}$ in topological field theory coupled to gravity.
(see e.g. \cite{Dij93} for a review)
As usual, we call the space spanned by $\{T^{\alpha,n}, n=0,1,2,\ldots\}$  the full
phase space  and  the subspace parameterized by $\{T^{\alpha,0}\}$ the small phase space
where  the indices $\alpha(=1,2, \ldots, N)$ and $n \geqs 0$  label
 the primary fields and  the level of gravitational descendants, respectively.
For a topological field theory, the most important quantities are correlation functions that
describe topological properties of the associated manifold.
The generating function of correlation functions is the full free energy defined by
${\mathcal F\/}(T)=\sum_{g=0}^\infty {\mathcal F\/}^{(g)}(T)=\sum_{g=0}^\infty
 \lan e^{\sum_{n,\alpha} T^{\alpha,n}\sigma_{n,\alpha}}\ran_g$
where $\lan\cdots\ran_g$ denotes the expectation value on a Riemann surface of
genus $g$ with respect to a classical action.  The so-called primary free energy $F(t)$ is just
   the genus-zero free energy\index{free energy!genus-zero} restricted
   on the small phase space, namely,  $ \mF^{(0)}|_{T^{\alpha,0}=t^\alpha, T^{n\geqs 1,\alpha}=0}=F(t)$.
After identifying the flat coordinates $t^\alpha$ with the genus-zero two-point functions
$\eta^{\alpha\beta}\pa^2{\mathcal F\/}^{(0)}/\pa T^{1,0}\pa T^{\beta,0}$, it turns out that
the hierarchy flows (\ref{bi-flow}) coincide with the genus-zero
topological recursion relation\cite{Wit90}.
\subsection{Models with Two Primary Fields}
\subsubsection{dToda hierarchy}
The two-dimensional Frobenius manifold  associated to the dToda is described
by the primary free energy\cite{EY94,Dub96}
\[
F_T(t)=\frac{1}{2}(t^1)^2t^2+e^{t^2},\quad t=(t^1,t^2),
\]
which satisfies $\mL_EF=2F$ with $E=t^1\pa_1+2\pa_2$.
The corresponding bi-hamiltonian structure can be deduced from the primary free energy as
\[
J_1^{\alpha\beta}=
\left(
\ba{cc}
0&\pa_x\\
\pa_x&0
\ea
\right),
\quad
J_2^{\alpha\beta}=
\left(
\ba{cc}
2e^{t^2}\pa_x+e^{t^2}t^2_x& t^1\pa_x\\
t^1\pa_x+t^1_x & 2\pa_x
\ea
\right),
\]
and the commuting hamiltonian flows of the dToda hierarchy are defined as (\ref{bi-flow})
with
\[
 \widetilde{H}_{\bet,n-1}=
  \frac{1}{n+\mu_\bet+1/2}\left(H_{\bet,n-1}-2\de_{1\beta}\frac{H_{2,n-2}}
  {(n+\mu_\beta+1/2)}\right),
\]
where $\mu_1=-1/2, \mu_2=1/2$ and the pair $(\beta,n)=(1,0)$ is resonant.
\subsubsection{Benney hierarchy}
The two-dimensional Frobenius manifold corresponding to the Benney hierarchy is described by
the primary free energy\cite{Dub96,CT01}
\[
F_B(t)=\frac{1}{2}(t^1)^2t^2+\frac{1}{2}(t^2)^2\left(\log t^2-\frac{3}{2}\right),
\]
which satisfies $\mL_EF=4F$ with $E=t^1\pa_1+2t^2\pa_2$. Just like the dToda hierarchy,
the associated bi-hamiltonian structure can be constructed as
\[
J_1^{\alpha\beta}=
\left(
\ba{cc}
0&\pa_x\\
\pa_x&0
\ea
\right),
\quad
J_2^{\alpha\beta}=
\left(
\ba{cc}
2\pa_x& t^1\pa_x+t^1_x\\
t^1\pa_x & 2t^2\pa_x+t^2_x
\ea
\right).
\]
The hamiltonian flows of the Benney hierarchy are defined as (\ref{bi-flow})
with
\[
  \widetilde{H}_{\bet,n-1}=
  \frac{1}{n+\mu_\bet+1/2}\left(H_{\bet,n-1}-2\de_{2\beta}\frac{H_{1,n-2}}
  {(n+\mu_\beta+1/2)}\right),
\]
where  $\mu_1=1/2, \mu_2=-1/2$ and the pair $(\beta,n)=(2,0)$ is resonant.
\subsubsection{dDym hierarchy}
For the dDym system, the associated two-dimensional Frobenius manifold is described by
the primary free energy\cite{Dub96,CT03a}
\[
F_D(t)=\frac{1}{2}(t^1)^2t^2-\frac{1}{2}\log t^2,
\]
which satisfies $\mL_EF=0$ with $E=t^1\pa_1-2t^2\pa_2$.
The corresponding bi-hamiltonian structure can be deduced from $F_D$ as
\[
J_1^{\alpha\beta}=
\left(
\ba{cc}
0&\pa_x\\
\pa_x&0
\ea
\right),
\quad
J_2^{\alpha\beta}=
\left(
\ba{cc}
\frac{2}{(t^2)^2}\pa_x-\frac{2}{(t^2)^3}t^2_x& t^1\pa_x-t^1_x\\
t^1\pa_x+2t^1_x & -2t^2\pa_x-t^2_x
\ea
\right),
\]
and the commuting hamiltonian flows of the dDym hierarchy are defined as (\ref{bi-flow})
with
\[
 \widetilde{H}_{\bet,n-1}=
  \frac{1}{n+\mu_\bet+1/2}\left(H_{\bet,n-1}+2\de_{1\beta}\frac{H_{2,n-4}}
  {(n+\mu_\beta+1/2)}\right),
\]
where $\mu_1=-3/2, \mu_2=3/2$ and the resonance occurs at the pair $(\beta,n)=(1,1)$.
\subsubsection{Polytropic gas dynamics}
Finally, we come to the Polytropic gas dynamics. The associated two-dimensional
Frobenius manifold is described by the primary free energy\cite{Dub96}
\[
F_h(t)=\frac{1}{2}(t^1)^2t^2+c_h(t^2)^{h+1},\quad h\neq -1,0,1,
\]
which is characterized by a polytropic exponent $h$ and satisfies $\mL_EF_h=(2+2/h)F_h$
with $E=t^1\pa_1+2h^{-1}t^2\pa_2$. The corresponding bi-hamiltonian structure
can be obtained as
\bean
&&J_1^{\alpha\beta}=
\left(
\ba{cc}
0&\pa_x\\
\pa_x&0
\ea
\right),\\
&&J_2^{\alpha\beta}=
\left(
\ba{cc}
2c_h(h^2-1)(t^2)^{h-1}\pa_x+c_h(h-1)(h^2-1)(t^2)^{h-2}t^2_x& t^1\pa_x+h^{-1}t^1_x\\
t^1\pa_x+(1-h^{-1})t^1_x & 2h^{-1}t^2\pa_x+h^{-1}t^2_x
\ea
\right),
\eean
and the commuting hamiltonian flows of the Polytropic gas dynamics are defined as (\ref{bi-flow})
with
\[
 \widetilde{H}_{\bet,n-1}=
  \frac{1}{n+\mu_\bet+1/2}\left(H_{\bet,n-1}-
  \sum_{k,\alpha}(R_k)^\alpha_\beta\frac{H_{\alpha,n-k-1}}
  {(n+\mu_\beta+1/2)}\right),
\]
where $\mu_1=1/h-1/2, \mu_2=1/2-1/h$ and the matrices $R_k$ are given by\cite{CLT07}
\[
(R_k)^\al_\bet=
\left\{
\ba{ll}
2(-1)^{m+1}\,\delta_{k,2m-1}\de_{\alpha,1}\de_{\beta,2}, & h=\frac{1}{m},
 \; m\geq 2,\\
 2(-1)^{m}\,\delta_{k,2m+1}\de_{\alpha,2}\de_{\beta,1}, & h=-\frac{1}{m},
\; m\geq 2, \\
0, & \mbox{otherwise}.
\ea\right.
\label{R}
\]
There are two cases to be discussed. (i) when $|h|^{-1}\not\in {\mathbb N\/}$ the bi-hamiltonian
recursive relation is just  Lenard's relation and the corresponding
Frobenius manifolds are nonresonant. (ii) when $h=1/m, |m|\in {\mathbb N\/}_{\geq 2}$ resonance
occurs at the pair $(2,m-1)$ for $m\geq 2$, whereas $(1,-m)$ for $m\leq -2$.

\section{Miura Transformations}
\subsection{WDVV Symmetries as Canonical Miura Maps}
Dubrovin \cite{Dub96} introduced two types of symmetries for the WDVV equations as follows.

Type 1. Legendre-type transformation $S_\kappa$:
\bea
\hat{t}_\alpha&=&\pa_\alpha\pa_\kappa F(t),\no\\
\frac{\pa^2 \hat{F}(\hat{t})}{\pa\hat{t}^\alpha\pa\hat{t}^\beta}&=&
\frac{\pa^2 F(t)}{\pa t^\alpha\pa t^\beta},
\label{S}\\
\hat{\eta}_{\alpha\beta}&=&\eta_{\alpha\beta},\no
\eea
where $\kappa=2,\cdots,N$ since $S_1$ is the identity transformation.
The Euler vector field transforms as
$\hat{E}(\hat{t})=E(t)$ with $\hat{d}=d-2+2d_\kappa$ and $\hat{\mu}=\mu$.
Based on above, it can be shown \cite{Dub96} that the vector field $\pa/\pa \hat{t}^\kappa$ should
be identified as the identity $e$, and  we have to interchange the indices $1$ and $\kappa$
after the transformation $S_\kappa$.

Type 2. Inversion transformation I:
\bea
\hat{t}^1&=&\frac{1}{2}\frac{t_\sig t^\sig}{t^N},\quad
\hat{t}^\alpha=\frac{t^\alpha}{t^N},(\alpha\neq 1,N),\quad
\hat{t}^N=-\frac{1}{t^N},\no\\
\hat{F}(\hat{t})&=&(\hat{t}^N)^2F(t)+\frac{1}{2}\hat{t}^1\hat{t}_\sig\hat{t}^\sig,
\label{I}\\
\hat{\eta}_{\alpha\beta}&=&\eta_{\alpha\beta}.\no
\eea
From above it can be shown that
\be
\hat{c}_{\alpha\beta\gamma}(\hat{t})=(t^N)^{-2}\frac{\pa t^\la}{\pa \hat{t}^\alpha}
\frac{\pa t^\mu}{\pa \hat{t}^\beta}\frac{\pa t^\nu}{\pa \hat{t}^\gamma}c_{\la\mu\nu}(t),
\label{cc}
\ee
and the Euler vector field transforms as $\hat{E}(\hat{t})=E(t)$ with $\hat{d}=2-d$
and  $\hat{\mu}=(d-1)(E_{11}-E_{NN})+\mu$.
Note that the Legendre-type transformation $S_\kappa$ is an involution transformations,
i.e., $(S_\kappa)^2=id$, while $I$ is  an involution transformations up to an equivalence
\[
I^2: (t^1, t^2,\cdots, t^{N-1}, t^N)\mapsto (t^1, -t^2,\cdots, -t^{N-1}, t^N).
\]

Below, we would like to investigate the canonical property of the WDVV symmetries
generated by $S_\kappa$ and $I$. In terms of flat coordinate $t^\alpha$,
 $\ga^{\alpha\beta}_{\sigma}(t)=0$ and the  WDVV transformations $S_\kappa$ and $I$,
 by virtue of (\ref{S}) and (\ref{I}), preserve the first hamiltonian
 structure, i.e.,
\[
J_1^{\alpha\beta}(t)\mapsto\hat{J}_1^{\alpha\beta}(\hat{t})=\eta^{\alpha\beta}\pa_x.
\]
For the second structure, since the flat metric $g^{\alpha\beta}(t)$ depends on $t$ nontrivially,
it would be interesting to work out the transformations of  second hamiltonian
structures under $S_\kappa$ and $I$.

Type 1. Legendre-type transformation $S_\kappa$.\\
From (\ref{sec}) and (\ref{Ga-c}), we have
\bea
g^{\alpha\beta}(t)&\stackrel{S_\kappa}{\mapsto}&\hat{g}^{\alpha\beta}(\hat{t})=
\hat{E}^\gamma(\hat{t})\hat{c}^{\alpha\beta}_\gamma(\hat{t})=g^{\alpha\beta}(t),
\label{S1}\\
\Gamma^{\alpha\beta}_\gamma(t) t^\gamma_x&\stackrel{S_\kappa}{\mapsto}&
\hat{\Gamma}^{\alpha\beta}_\gamma(\hat{t})\hat{t}^\gamma_x=
\hat{c}^{\alpha\ep}_\gamma(\hat{t})\hat{t}^\gamma_x\left(\frac{1}{2}-\hat{\mu}\right)_\ep^\beta
=\Gamma^{\alpha\beta}_\gamma(t)t^\ga_x,
\label{S2}
\eea
where $\hat{\Gamma}^{\alpha\beta}_\gamma(\hat{t})$ is the connection with respect to
the metric $\hat{g}^{\alpha\beta}(\hat{t})$. Therefore,
\[
J_2^{\alpha\beta}(t)\stackrel{S_\kappa}{\mapsto}\hat{J}_2^{\alpha\beta}(\hat{t})=
J_2^{\alpha\beta}(t),
\]
which is just a coordinate transformation.

Type 2. Inversion transformation I.\\
In this case, using (\ref{sec}), (\ref{Ga-c}) and (\ref{cc}), we have
\bea
g^{\alpha\beta}(t)&\stackrel{I}{\mapsto}&
\hat{g}^{\alpha\beta}(\hat{t})
=(t^N)^{-2}\frac{\pa t_\la}{\pa \hat{t}_\alpha}\frac{\pa t_\mu}{\pa \hat{t}_\beta}g^{\la\mu}(t),
\label{I1}\\
\Gamma^{\alpha\beta}_\gamma(t) t^\gamma_x&\stackrel{I}{\mapsto}&
\hat{\Gamma}^{\alpha\beta}_\gamma(\hat{t})\hat{t}^\gamma_x
=(t^N)^{-2}\frac{\pa t_\la}{\pa \hat{t}_\alpha}
\frac{\pa t_\mu}{\pa \hat{t}_\ep}c^{\la\mu}_\nu(t) t^\nu_x
\left(\frac{1}{2}-\mu+(1-d)(E_{11}-E_{NN})\right)_\ep^\beta,
\label{I2}
\eea
and hence
\[
J_2^{\alpha\beta}(t)\stackrel{I}{\mapsto} \hat{J}_2^{\alpha\beta}(\hat{t})=
(t^N)^{-2}\frac{\pa t_\la}{\pa \hat{t}_\alpha}\frac{\pa t_\mu}{\pa \hat{t}_\beta}
\left[g^{\la\mu}(t)\pa_x+c^{\la\mu}_\nu(t) t^\nu_x\left(\frac{1}{2}-\mu_\beta+
(1-d)(\de_{\beta,1}-\de_{\beta,N})\right) \right].
\]
Since $ \hat{J}_1^{\alpha\beta}(\hat{t})$ and $ \hat{J}_2^{\alpha\beta}(\hat{t})$
form a flat pencil with respect to the new primary free energy $\hat{F}(\hat{t})$ and hence
symmetries of WDVV can be viewed as canonical Miura transformations between
 bi-hamiltonian system of hydrodynamic type.
\subsection{Miura Transformations between Two-Primary Models}
Now let us apply WDVV transformations on those two-primary models $(N=2)$ listed
in Table 1. Here we have $(S_2)^2=I^2=id$.
Note that we have to interchange the indices $1$ and $2$ after the transformation $S_2$.
\subsubsection{Benney $\leftrightarrow$ dToda}
Under the Legendre-type transformation $S_2$ the Benney system
transforms as\cite{Dub96}
\[
\left\{
\ba{ccc}
F_B(t)&=&\frac{1}{2}(t^1)^2t^2+\frac{1}{2}(t^2)^2(\log t^2-\frac{3}{2})\\
E_B(t)&=&t^1\frac{\pa}{\pa t^1}+2t^2\frac{\pa}{\pa t^2}\\
(J_1, J_2)_B&&
\ea
\right\}\stackrel{S_2}{\leftrightarrow}
\left\{
\ba{ccc}
\hat{F}_T(\hat{t})&=&\frac{1}{2}(\hat{t}^1)^2\hat{t}^2+e^{\hat{t}^2}\\
\hat{E}_T(\hat{t})&=&\hat{t}^1\frac{\pa}{\pa \hat{t}^1}+2\frac{\pa}{\pa \hat{t}^2}\\
(\hat{J}_1, \hat{J}_2)_T&&
\ea
\right\}
\label{BT}
\]
where
\[
t^1=\hat{t}^2,\quad t^2=e^{\hat{t}^1}.
\]
Let us verify the canonical property of $S_2$ for $J_2$.
From (\ref{S1}) and (\ref{S2}) we have
\bean
g(t)&\stackrel{S_2}{\mapsto}&\hat{g}(\hat{t})=g(t)
=
\left(
\ba{cc}
2 & \hat{t}^2\\
\hat{t}^2 & 2e^{\hat{t}^1}
\ea
\right)
\stackrel{(1\leftrightarrow 2)}{=}
\left(
\ba{cc}
2e^{\hat{t}^2} & \hat{t}^1\\
\hat{t}^1 & 2
\ea
\right),\\
\Gamma^{\alpha\beta}_\gamma(t) t^\gamma_x&\stackrel{S_2}{\mapsto}&
\hat{\Gamma}^{\alpha\beta}_\gamma(\hat{t})\hat{t}^\gamma_x
=\Gamma^{\alpha\beta}_\gamma(t) t^\gamma_x=
\left(
\ba{cc}
0 & \hat{t}^2_x\\
0 & \hat{t}^1_xe^{\hat{t}^1}
\ea
\right)
\stackrel{(1\leftrightarrow 2)}{=}
\left(
\ba{cc}
\hat{t}^2_xe^{\hat{t}^2} & 0\\
\hat{t}^1_x & 0
\ea
\right).
\eean
Therefore, the bi-hamiltonian structures of the dToda and Benney hierarchies
are canonically related under the transformation $S_2$.
\subsubsection{Benney $\leftrightarrow$ dDym}
Under the inversion transformation $I$ the Benney system
transforms as\cite{Dub96}
\[
\left\{
\ba{ccc}
F_B(t)&=&\frac{1}{2}(t^1)^2t^2+\frac{1}{2}(t^2)^2\log t^2\\
E_B(t)&=&t^1\frac{\pa}{\pa t^1}+2t^2\frac{\pa}{\pa t^2}\\
(J_1, J_2)_B&&
\ea
\right\}\stackrel{I}{\leftrightarrow}
\left\{
\ba{ccc}
\hat{F}_D(\hat{t})&=&\frac{1}{2}(\hat{t}^1)^2\hat{t}^2-\frac{1}{2}\log \hat{t}^2\\
\hat{E}_D(\hat{t})&=&\hat{t}^1\frac{\pa}{\pa \hat{t}^1}-2\hat{t}^2\frac{\pa}{\pa \hat{t}^2}\\
(\hat{J}_1, \hat{J}_2)_D&&
\ea
\right\}
\label{BD}
\]
where
\[
t^1=\hat{t}^1,\quad t^2=-\frac{1}{\hat{t}^2}.
\]
To verify the canonical property under the inversion $I$, let us define the matrix
\[
S^\alpha_{\;\;\la}=\left(\frac{\pa t_\la}{\pa\hat{t}_\alpha}\right)=
\left(
\ba{cc}
\frac{1}{(\hat{t}^2)^2}& 0\\
0&1
\ea
\right),
\]
then from (\ref{I1}) and (\ref{I2}) we have
\bean
g^{\alpha\beta}(t)&\stackrel{I}{\mapsto}&
\hat{g}^{\alpha\beta}(\hat{t})=\frac{1}{(t^2)^2}S^\alpha_{\;\;\la}g^{\la\mu}(S^t)_\mu^{\;\;\beta}
=
\left(
\ba{cc}
\frac{2}{(\hat{t}^2)^2}& \hat{t}^1\\
\hat{t}^1& -2\hat{t}^2
\ea
\right),\\
\Gamma^{\alpha\beta}_\gamma(t) t^\gamma_x&\stackrel{I}{\mapsto}&
\hat{\Gamma}^{\alpha\beta}_\gamma(\hat{t})\hat{t}^\gamma_x
=(t^2)^{-2}S^\alpha_{\;\;\la}(c^{\la\mu}_\nu(t) t^\nu_x)(S^t)_\mu^{\;\;\ep}
\left(\frac{1}{2}+2(E_{11}-E_{22})-\mu\right)_\ep^\beta,\\
&&\qquad \qquad =
\left(
\ba{cc}
-\frac{2\hat{t}^2_x}{(\hat{t}^2)^3}& -\hat{t}^1_x\\
2\hat{t}^1_x& -\hat{t}^2_x
\ea
\right).
\eean
Hence, the bi-hamiltonian structure of the Benney hierarchy is mapped to that of the dDym hierarchy.
\subsubsection{polytropic gas $\leftrightarrow$ dDym}
Under the Legendre-type transformation $S_2$ the Polytropic gas dynamics
with $h= 1/2$ transforms as
\[
\left\{
\ba{ccc}
F_{\frac{1}{2}}(t)&=&\frac{1}{2}(t^1)^2t^2+c_{1/2}(t^2)^{3/2}\\
E_{\frac{1}{2}}(t)&=&t^1\frac{\pa}{\pa t^1}+4t^2\frac{\pa}{\pa t^2}\\
(J_1, J_2)_{\frac{1}{2}}&&
\ea
\right\}\stackrel{S_2}{\leftrightarrow}
\left\{
\ba{ccc}
\hat{F}_D(\hat{t})&=&\frac{1}{2}(\hat{t}^1)^2\hat{t}^2-\frac{1}{2}\log \hat{t}^2\\
\hat{E}_D(\hat{t})&=&\hat{t}^1\frac{\pa}{\pa \hat{t}^1}-2\hat{t}^2\frac{\pa}{\pa \hat{t}^2}\\
(\hat{J}_1, \hat{J}_2)_D
\ea
\right\}
\label{hD}
\]
where
\[
t^1=\hat{t}^2,\quad t^2=\frac{1}{2(\hat{t}^1)^2},\quad c_{1/2}=\frac{2^{3/2}}{3}.
\]
\subsubsection{polytropic gas $\leftrightarrow$ polytropic gas}
Under the Legendre-type transformation $S_2$, the polytropic gas system
with $h\neq 1/2$ transforms as
\[
\left\{
\ba{ccc}
F_{h\neq 1/2}(t)&=&\frac{1}{2}(t^1)^2t^2+c_h(t^2)^{1+h}\\
E_{h\neq 1/2}(t)&=&t^1\frac{\pa}{\pa t^1}+\frac{2}{h}t^2\frac{\pa}{\pa t^2}\\
(J_1, J_2)_{h\neq 1/2}&&
\ea
\right\}\stackrel{S_2}{\leftrightarrow}
\left\{
\ba{ccc}
\hat{F}_{\frac{h}{h-1}}(\hat{t})&=&\frac{1}{2}(\hat{t}^1)^2\hat{t}^2+\hat{c}_{\frac{h}{h-1}}
(\hat{t}^2)^{1+\frac{h}{h-1}}\\
\hat{E}_{\frac{h}{h-1}}(\hat{t})&=&\hat{t}^1\frac{\pa}{\pa \hat{t}^1}-
\frac{2(1-h)}{h}\hat{t}^2\frac{\pa}{\pa \hat{t}^2}\\
(\hat{J}_1, \hat{J}_2)_{\frac{h}{h-1}}&&
\ea
\right\}
\label{hSh}
\]
where
\[
t^1=\hat{t}^2,\quad t^2=\left(\frac{\hat{t}^1}{c_hh(1+h)}\right)^{\frac{1}{h-1}},\quad
\hat{c}_{\frac{h}{h-1}}=\frac{(h-1)^2}{h(2h-1)}(c_hh(h+1))^{\frac{1}{1-h}}.
\]
On the other hand, under the inversion transformation $I$ the polytropic gas system
transforms as
\[
\left\{
\ba{ccc}
F_h(t)&=&\frac{1}{2}(t^1)^2t^2+c_h(t^2)^{1+h}\\
E_h(t)&=&t^1\frac{\pa}{\pa t^1}+\frac{2}{h}t^2\frac{\pa}{\pa t^2}\\
(J_1, J_2)_h&&
\ea
\right\}\stackrel{I}{\leftrightarrow}
\left\{
\ba{ccc}
\hat{F}_{-h}(\hat{t})&=&\frac{1}{2}(\hat{t}^1)^2\hat{t}^2+\hat{c}_{-h}(\hat{t}^2)^{1-h}\\
\hat{E}_{-h}(\hat{t})&=&\hat{t}^1\frac{\pa}{\pa \hat{t}^1}-\frac{2}{h}\hat{t}^2
\frac{\pa}{\pa \hat{t}^2}\\
(\hat{J}_1, \hat{J}_2)_{-h}&&
\ea
\right\}
\label{hIh}
\]
where
\[
t^1=\hat{t}^1,\quad t^2=-\frac{1}{\hat{t}^2},\quad \hat{c}_{-h}=(-1)^{1-h}c_h.
\]
\subsection{Moduli Space}
Based on the above discussions, we find that the relationship between the four types of
two-primary models can be summarized by the following sequences of WDVV transformations:

{\bf Resonant sequence.\/}\\
For those two-dimensional Frobenius manifolds  including $F_T$, $F_B$, $F_D$ and
$F_h$ with $|h|^{-1}\in {\mathbb N\/}_{\geq 2}$, they are all resonant and the Miura
transformations between them constitute a sequence of WDVV transformations:
\be
F_0\cdots F_{\frac{1}{n+1}}\stackrel{S_2}{\leftrightarrow} F_{-\frac{1}{n}}
\stackrel{I}{\leftrightarrow}F_{\frac{1}{n}}\cdots
F_{-\frac{1}{2}}\stackrel{I}{\leftrightarrow}
 F_{\frac{1}{2}}\stackrel{S_2}{\leftrightarrow}F_D
\stackrel{I}{\leftrightarrow}F_B\stackrel{S_2}{\leftrightarrow}
F_T,
\label{seq}
\ee
where the coefficient $c_{\frac{1}{n}} (n\geq 2)$ of $F_{\frac{1}{n}}$ can be computed as
$c_{\frac{1}{n}}=(-1)^nn^{\frac{n-1}{n}}(n!)^{\frac{2}{n}}/(n^2-1).
$
Since $F_0$ is an invariant function under $S_2$ and $I$ transformations, thus, on the left end,
the resonant sequence terminates at the limiting function $F_0$.
On the right end, the resonant sequence terminates at
$F_T$ because  under inversion transformation $I$ the transformed primary free energy no longer
satisfies the quasi-homogeneity condition (\ref{quasi}).

{\bf Nonresonant sequences.\/}\\
For those polytropic gas systems defined by  $F_h$ with $|h|^{-1}\not\in {\mathbb N\/}$, we have
\be
F_0\cdots\stackrel{I}{\leftrightarrow}F_{\frac{h}{kh+1}}\stackrel{S_2}{\leftrightarrow}
\cdots F_{\frac{h}{h+1}}\stackrel{S_2}{\leftrightarrow}F_{-h}\stackrel{I}{\leftrightarrow}
 F_{h}\stackrel{S_2}{\leftrightarrow}F_{\frac{h}{h-1}}\cdots
 \stackrel{S_2}{\leftrightarrow}F_{\frac{h}{kh-1}}\stackrel{I}{\leftrightarrow}\cdots F_{0}.
 \label{seq2}
\ee
It may be noticed that $F_2$ is also an invariant function of $S_2$, thus the sequence
 generated by $F_2$ is semi-infinite; namely,
\[
F_0\cdots\stackrel{I}{\leftrightarrow}F_{\frac{2}{2k+1}}\stackrel{S_2}{\leftrightarrow}
\cdots F_{\frac{2}{3}}\stackrel{S_2}{\leftrightarrow}F_{-2}\stackrel{I}{\leftrightarrow}
 \stackrel{\stackrel{S_2}{\curvearrowright}}{F_{2}}.
\]
There are some peculiar properties associated with nonresonant-sequences.

(i) {\it No two polytropic exponents are the same in a sequence of WDVV transformations\/}.
It can be seen from the fact that for a nonresonant-sequence (\ref{seq2}) containing
 $F_h$, the possible values of polytropic exponents are $\pm h/(kh\pm 1)$,
 where $k\geq 0$, $h>0$ and $h^{-1}\not\in {\mathbb N\/}$.
If two exponents in a sequence coincide, it must be one of the  cases
$h/(kh+1)=\pm h/(k'h\pm 1)$. However, they all give  contradictions.

(ii) {\it Two nonresonant sequences either have no common elements or overlap completely\/}.
If two nonresonant sequences associated with $F_h$ and $F_{h'}$ ($h\neq h'$, $h, h'>0$)
have a common element, then we have $\pm h/(kh\pm 1)=\pm h'/(k'h'\pm 1)$
where $k, k'\geq 0$, and $h$ and $h'$ belong to two different sequences.
However, for example, if $h/(kh+1)=h'/(k'h'+1)$ which implies that
$h'=h/((k-k')h+1)$ and thus $h$ and $h'$ belong to the same sequence. Other cases
can be verified in a similar manner.

The above properties motivate us to introduce a fundamental interval parameterized
 by the polytropic exponent $h$ so that every sequence has a representative value of $h$
 in this interval. It turns out that the moduli space  for two-primary models
can be chosen as one of the following intervals:
\[
{\mathbf  D\/}_n=\left\{h\left| h\in  \left[\frac{2}{2n+1}, \frac{1}{n}\right]\right.\right\},\quad
n=0,1,2,\cdots
\]
with the proviso that an ``effective" polytropic exponent was assigned to dToda, Benney, and dDym
as $\infty$, $1$, and $-1$, respectively.
Note that ${\mathbf D\/}_n$ are isomorphic to the quotient space
${\mathbb R\/}P_1^\times/\{S_2,I\}$ where
${\mathbb R\/}P_1^\times={\mathbb R\/}\cup  \{\infty\}\setminus\{0\}$, and
are isomorphic to each other via the map $S_2I: {\mathbf  D\/}_n \rightarrow
{\mathbf  D\/}_{n+1}$.
It was pointed out \cite{Dub96} that a Frobenius manifold is called {\it reduced\/}
if it satisfies the inequality:
\be
0\leq q_\alpha\leq d\leq 1,
\label{red}
\ee
and one can reduce by the transformations $S_\kappa$ and
$I$ any solution of WDVV to a solution with property (\ref{red}).
It is indeed the case for two-primary models since
${\mathbf D\/}_0=\{h|h\in [2,\infty]\}$ is just the only fundamental interval satisfying
 the  inequality (\ref{red}).

\section{WDVV Symmetries as Quantum Miura Maps}
So far we only focus our discussions on bi-hamiltonian systems at  genus-zero level.
For higher genus we may consult the work of Dubrovin and Zhang (DZ)\cite{DZ98},
which  provide us a starting point to extend a bi-hamiltonian hydrodynamic system to its
dispersive counterpart and obtain the corresponding commuting flows up to genus-one corrections.
The DZ approach \cite{DZ98} consists  two main ingredients:
(a) introducing slow spatial and time variables scaling
$ T^{\al,n}\rightarrow \epsilon T^{\al,n}, n=0,1,2,\ldots$. (b)  changing the
full free energy as $ \mF\rightarrow \sum_{g=0}^{\infty}\epsilon^{2g-2}\mF^{(g)}$,
where $\epsilon$ is the parameter of genus expansion. Thus all of the corrections become
series in $\epsilon$.
To get an unambiguous genus-one correction of the hamiltonian flows (\ref{bi-flow})
one may expand the flat coordinates up to the $\ep^2$ order as
$ t_{\al}=t_{\al}^{(0)}+\epsilon^2t_{\al}^{(1)}+O(\epsilon^4), t_{\al}=\eta_{\al\beta}t^{\beta}$,
 where $t_{\al}^{(0)}$ are the ordinary flat coordinates,
and $t_{\al}^{(1)}$ are the genus-one correction
defined by $t_{\al}^{(1)}=\pa^2\mF^{(1)}(T)/\pa T^{\al,0}\pa x$.
The genus-one part of the free energy has the form\cite{DZ98},
\[
 \mF^{(1)}(T) = \frac{1}{24}\log \det M^{\al}_{\beta}(t,\pa_xt)+G(t),
\]
where the matrix $ M^{\al}_{\beta}(t,\pa_xt) = c^{\al}_{\beta\ga}(t)t^{\ga}_x$
and $G(t)$ is the so-called $G$-function which depends on $(t^2,\cdots,t^N)$ and
satisfies the Getzler's equation \cite{Get97}.
Using $c^{\al}_{\beta\ga}(t)$ and $\mF^{(1)}(t)$
and consulting the procedure developed  in \cite{DZ98}(c.f. Theorem 1 and 2 and Proposition 3),
one can obtain the genus-one corrections of the Poisson brackets and hierarchy flows.
This means that the bi-hamiltonian structure $J_1$ and $J_2$ and the
hamiltonians will receive corrections up to $\ep^2$ such that the hamiltonian flows
still commute with each other. From (\ref{S}) and (\ref{I}), we have
\bean
&&M^\alpha_\beta(t)\stackrel{S_\kappa}{\mapsto}\hat{M}^\alpha_\beta(\hat{t})
=M^\alpha_\beta(t),\\
&&M^\alpha_\beta(t)\stackrel{I}{\mapsto}\hat{M}^\alpha_\beta(\hat{t})
=(t^N)^{-2}\frac{\pa t_\lambda}{\pa \hat{t}_\alpha}\frac{\pa t^\mu}
{\pa \hat{t}^\beta}M^\la_\mu(t).
\eean
On the other hand, the transformations of the $G$-function under WDVV symmetries have
been obtained by Strachan as\cite{Str03}
\bean
&&G(t)\stackrel{S_\kappa}{\mapsto}\hat{G}(\hat{t})=G(t)-\frac{1}{24}\log
\det\left(\frac{\pa \hat{t}^\alpha}{\pa t^\beta}\right),\\
&&G(t)\stackrel{I}{\mapsto}\hat{G}(\hat{t})=G(t)+\left(\frac{N}{24}-\frac{1}{2}\right)\log t^N.
\eean
Combining the above results we obtain the WDVV transformations for
the genus-one free energy $\mF^{(1)}$ as
\bean
&&\mF^{(1)}(t)\stackrel{S_\kappa}{\mapsto}\hat{\mF}^{(1)}(\hat{t})
=\mF^{(1)}(t)-\frac{1}{24}\log \det c_{\kappa\beta}^\alpha(t),\\
&&\mF^{(1)}(t)\stackrel{I}{\mapsto}\hat{\mF}^{(1)}(\hat{t})
=\mF^{(1)}(t)+\left(\frac{N}{24}-\frac{1}{2}\right)\log t^N,
\eean
where we have used the formula $\det\left(\frac{\pa \hat{t}^\alpha}{\pa t^\beta}\right)
=(t^N)^{-N}$ for the inversion transformation $I$.

For the two-primary models ($N=2$),  we list the $G$-function and the genus-one free energy
$\mF^{(1)}$  in Table \ref{FG2}.
\begin{center}
\begin{threeparttable}
\caption{$G$-function and genus-one free energy $\mF^{(1)}$ of two-primary models}
\label{FG2}
\index{free energy}
\index{G-f@$G$-function}
\begin{tabular*}{135mm}[]{ccc@{}c}
\toprule
  model  &      $G$-function   & $\mF^{(1)}$  \\
\midrule
  polytropic gas  &   $-\frac{1}{24}\frac{(2-h)(3-h)}{h}\log t^2$
    & $\frac{1}{24}
   \log\left[(t^1_x)^2-h(h^2-1)c_h(t^2)^{h-2}(t^2_x)^2\right]$\\
   & &  $\qquad\quad-\frac{1}{24}\frac{(h-2)(h-3)}{h}\log t^2$
   \\ \\
    dDym  &  $\frac{1}{2}\log t^2$ &
  $\frac{1}{24}\log\left[(t^2)^3(t^1_x)^2+(t^2_x)^2\right] + \frac{3}{8}\log t^2$
   \\ \\
  Benney &  $-\frac{1}{12}\log t^2$ &
  $\frac{1}{24}\log\left[t^2(t^1_x)^2-(t^2_x)^2\right] - \frac{1}{8}\log t^2$
   \\ \\
   dToda  &    $-\frac{1}{24}t^2$  &
  $\frac{1}{24}\log\left[(t^1_x)^2-(t^2_x)^2e^{t^2}\right] - \frac{1}{24} t^2$
   \\
\bottomrule
\end{tabular*}
\end{threeparttable}
\end{center}

In fact, both resonant and nonresonant sequences of WDVV transformations
for genus-zero free energy  can be promoted to genus-one level.
For example, consider the genus-one free energy of the Benney system under the
transformation  $S_2$:
\bean
\mF_B^{(1)}(t)\stackrel{S_2}{\mapsto}\hat{\mF}^{(1)}(\hat{t})
&=&\frac{1}{24}\log\left[(\hat{t}^2_x)^2-(\hat{t}^1_x)^2e^{\hat{t}^1}\right]-\frac{1}{24}\hat{t}^1,\\
&\stackrel{(1\leftrightarrow 2)}{=}&
\frac{1}{24}\log\left[(\hat{t}^1_x)^2-(\hat{t}^2_x)^2e^{\hat{t}^2}\right]-\frac{1}{24}\hat{t}^2,
\eean
which is just the genus-one free energy  of the dToda hierarchy,
$\hat{\mF}_T^{(1)}(\hat{t})$. In a similar manner,
other transformations between genus-zero free energy in the sequences (\ref{seq}) and (\ref{seq2})
can be lifted to genus-one free energy without difficulty.
Therefore, it seems to suggest that the bi-hamiltonian structures
of the two-primary models, up to the genus-one corrections, are canonically related
under WDVV transformations. A direct verification of this claim remains to be worked out.

{\bf Acknowledgement\/}\\
This work is supported by the National Science Council of Taiwan
under Grant numbers NSC95-2112-M-194-005-MY2 and NSC97-2112-M-194-002-MY3.


\end{document}